# Electron Ptychography Images Hydrogen Atom Superlattices and 3D Inhomogeneities in Palladium Hydride Nanoparticles

*Zixiao Shi[1], Qihao Li[1], Himani Mishra[2], Desheng Ma[2], Héctor D. Abruña[1*], David A. Muller[2,3*]*

[1] Department of Chemistry and Chemical Biology, Cornell University, Ithaca, NY, 14853

[2] School of Applied & Engineering Physics, Cornell University, Ithaca, NY, 14853

[3] Kavli Institute at Cornell for Nanoscale Science, Cornell University, Ithaca, NY, 14853

Email: dm24@cornell.edu, hda1@cornell.edu

# Electron Ptychography Images Hydrogen Atom Superlattices and 3D Inhomogeneities in Palladium Hydride Nanoparticles

**ABSTRACT**

When hydrogen atoms occupy interstitial sites in metal lattices, they form metal hydrides ($MH_x$), whose structural and electronic properties can differ significantly from the host metals. Owing to the small size of hydrogen atom and its unique interactions with the host metal, $MH_x$ is of broad interest in both fundamental science and technological applications. Determining where the hydrogen is located within the $MH_x$, and whether it orders on the partially occupied interstitial sites is crucial for predicting and understanding the resultant physical and electronic properties of the hydride. Directly imaging hydrogen within a host material remains a major challenge due to its weak interaction with X-rays and electrons in conventional imaging techniques. Here, we employ electron ptychography, a scanning transmission electron microscopy technique, to image the three-dimensional (3D) distribution of H atoms in Palladium hydrides ($PdH_x$) nanocubes, one of the most studied and industrially relevant $MH_x$ materials. We observe an unexpected one-dimensional superlattice ordering of hydrogen within the $PdH_x$ nanocubes and 3D hydrogen clustering in localized regions within $PdH_x$ nanocubes, revealing spatial heterogeneity in metal hydride nanoparticles previously inaccessible by other methods.

**INTRODUCTION**

Hydrogen -- the lightest, smallest, and most abundant chemical element in the universe plays a foundational role in chemistry and physics.[1,2] Yet despite its apparent simplicity, hydrogen continues to pose major scientific challenges, notably the difficulty of imaging hydrogen atoms in metal hydrides ($MH_x$).[3] Due to its low scattering cross-section and much smaller electrostatic potential relative to metal atoms, hydrogen often remains invisible in conventional X-ray and electron imaging.[4] However, resolving the atomic-scale distribution of hydrogen atoms is crucial, as it significantly influences the electronic, magnetic, thermal, and electrochemical properties of $MH_x$, distinguishing them from their pure metal counterparts.[5–8] The unique characteristics of hydrogen, including its role as the smallest possible anion and its ability to modulate host lattice properties, makes $MH_x$ a valuable system in chemistry and physics, particularly superconductivity.[9] Beyond their scientific relevance, $MH_x$ are also essential for practical applications, including hydrogen storage, sensing technologies, and electrochemical energy systems.[6,7,10,11]

Palladium hydride ($PdH_x$), first discovered in 1866 by the chemist Thomas Graham, remains one of the most studied metal hydrides today.[12,13] $PdH_x$ exhibits unusual properties, including an inverse isotope effect, an anomaly in specific heat near 50 K, and one of the highest superconducting transition temperatures among $MH_x$ under ambient pressure.[14,15] It is also widely studied for its tunable hydrogen absorption behavior, which is relevant for studies in hydrogen storage, plasmonic and electrochemistry.[16–20] Many of these intriguing phenomena are governed by the interplay between electronic and phonon band structures, which in turn depend sensitively on the crystal structure, particularly the precise geometric arrangement of hydrogen atoms.[15]

Although imaging hydrogen atoms has long been of great interest, conventional techniques lack the resolution needed to pinpoint their positions at the atomic scale. X-ray and neutron imaging cannot yet achieve atomic resolution for hydrogen positions.[21,22] Atom probe tomography provides 3D hydrogen distribution but with nanometer-scale resolution and limited applicability to systems like nanoparticles.[23] Scanning transmission electron microscope (STEM) imaging methods, such as scanning electron nanobeam diffraction (SEND) and electron energy loss spectroscopy (EELS) can probe hydrogen uptake at nanometer resolution.[24,25] Other STEM imaging techniques such as annular bright field (ABF) and integrated differential phase contrast (iDPC), have shown the potential to visualize hydrogen atoms.[26–28] However, they do not solve how electrons scatter within the sample, relying instead on simplified interpretations of scattered signals. As a result, they require ultrathin samples of hydrides to avoid multiple scattering artifacts, and are highly sensitive to sample tilt, often leading to inaccurate assignments of hydrogen positions.[29,30] The core difficulty is that hydrogen atoms scatter electrons so weakly that their signal is easily overwhelmed by the surrounding metal background and electron probe tails.

Multislice Electron Ptychography (MEP) has been successfully utilized for imaging light elements such as lithium (Li), nitrogen (N), and oxygen (O), in metal compounds by solving electron multiple scattering within the sample.[31–34] Here, we extend this approach to directly resolve 3D distribution of H atoms within $PdH_x$ (x = 0.4) nanoparticles.[35] We note also useful contemporary work in imaging H in bulk $TiH_x$ and $NbH_x$, where a systematic comparison between MEP, iDPC, and ABF is made, again highlighting some of the challenges discussed above.[36]

Here, we observe that some hydrogen atoms in $PdH_x$ arrange into repeating ordering, a superlattice along [1$\bar{1}$1] direction, where every 2$^{nd}$ plane of the octahedral interstitial sites is occupied. The occupancy of every 2$^{nd}$ site is useful for testing and ruling out the effect of electron-probe-tail artifacts in imaging. This superlattice ordering is somewhat unexpected, as at room temperature, hydrogen can diffuse rapidly in Pd at low concentrations.[37] However, the strong lattice distortions induced by the hydrogen atoms also lead to self-trapping, which suggests local clustering may be possible.[37] Long-range ordering of hydrogen superlattices with different orientations has been reported at low temperatures, but with considerable variability.[38–40] In our case, the superlattice domain sizes might be too small to detect by bulk diffraction methods and appear to be concentrated near the surface regions of the nanoparticles. Consistent with this, our MEP images show a 3D clustering of hydrogen atoms forming ordered domains within the $PdH_x$ nanocubes. These results demonstrate the utility of electron ptychography as a quantitative method for uncovering hidden hydrogen ordering in metal hydrides, enabling direct 3D mapping of hydrogen and offering new insights into their atomic structures.

DISCUSSION

**Ptychography images hydrogen atoms in 3D with 0.3 Å lateral resolution**

To visualize hydrogen atoms in metal hydrides, we chose palladium hydride ($PdH_x$) nanoparticles—a widely studied system where hydrogen incorporates into the crystal lattice of Pd.[41–44] In our case, $PdH_x$ nanocubes are made by a wet-chemistry route from Pd nanocubes (see SI experiments, Figure S1-S5).

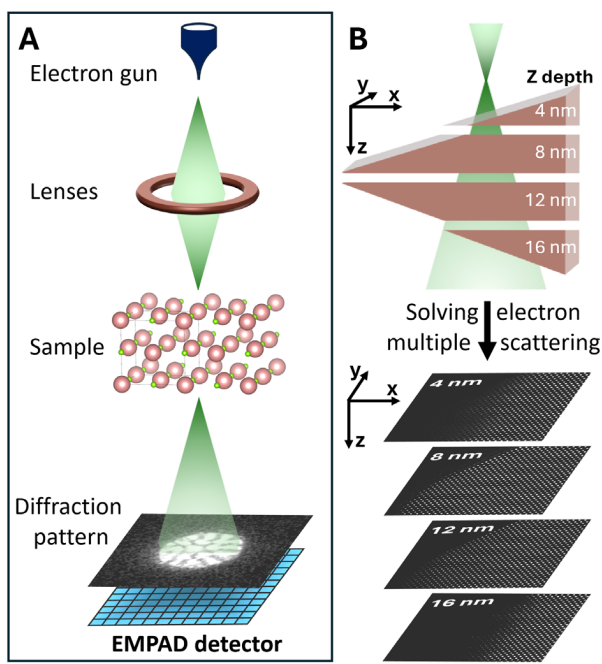

**Figure 1.** (A) Schematic of the ptychography experimental setup, using an EMPAD direct electron detector is used to collect the full diffraction pattern at every electron probe position (x, y). (B) Multislice electron ptychography calculates electron multiple scattering within the object, using the parallax of the diverging beam to recover depth information from an x-y scan with a single z-focus.

Multislice electron ptychography is a STEM imaging technique that reveals the 3D structure of materials by solving the inverse problem of recovering the 3D potential from an oversampled set of electron diffraction patterns, using a forward model that accurately models the electron beam propagation through a thick (i.e. multiply-scattering) sample.[34,45] When a focused electron beam hits the sample, it produces a diffraction pattern (Figure 1A). In electron ptychography, we collect the entire diffraction pattern at every probe position with an electron microscopy pixelated array detector (EMPAD) to construct a four-dimensional data set $P(x, y, p_x, p_y)$ of scattering probabilities in position and momentum space, and computationally work backward to determine what 3D structure of the material produces that pattern.[46,47]

This method accounts for the fact that electron beams can scatter multiple times as they pass through the sample, changing their shape on distances of one nanometer or less. By properly accounting for this multiple scattering, we can correct for beam spreading, channeling and mistilt, and we can then differentiate between electrons scattered from light atoms and those from heavy atoms; meanwhile, we obtain 3D information at different Z depths (Z axis along the electron beam, Figure 1B, Figure S3).

Electron ptychography enables direct imaging of hydrogen atoms in $PdH_x$ nanocubes, achieving 0.3 Å lateral resolution by using phase information from overlapping diffraction features (Figure S9-S11).

In the $PdH_x$ crystal lattice, hydrogen can occupy either octahedral or tetrahedral interstitial sites.[48,49] When viewed along the [110] crystallographic orientation, where these sites project into distinct positions relative to the surrounding Pd columns, we can distinguish between empty sites, octahedral H sites, and tetrahedral H sites.

Without any image contrast adjustment, both HAADF and ptychography clearly show the positions of Pd atoms, while ptychography provides much better lateral resolution, especially for thick regions (Figure 2A-B). By adjusting the contrast of both HAADF and ptychography images, without any additional filtering, we find that the ptychography images reveal the far-weakly-scattering hydrogen atoms occupying octahedral interstitial sites (Figure 2C). In contrast, conventional methods like HAADF do not detect any hydrogen signal, and iDPC images either fail to reveal hydrogen or indicate incorrect positions (Figure S6-S7). Finally, by directly comparing ptychography images of pure Pd and PdH$_x$ nanocubes (Figure S8), we confirm that hydrogen atoms appear exclusively at the octahedral sites, with no evidence of occupancy at tetrahedral sites in our PdH$_x$ nanocubes.[48,49]

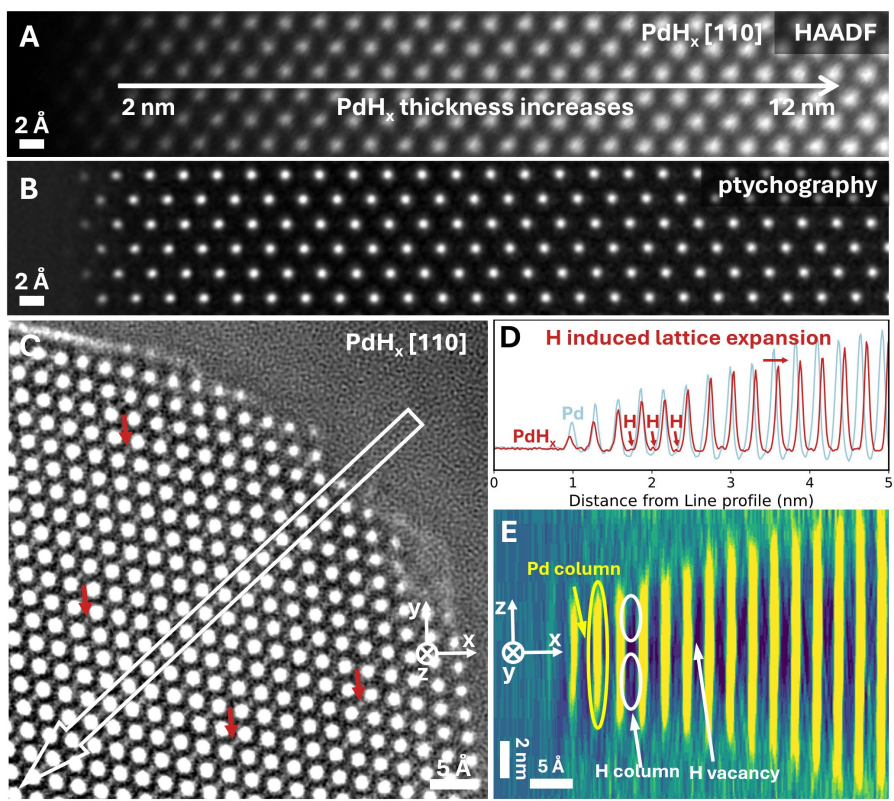

**Figure 2.** (A–B) HAADF and ptychography images of PdH$_x$ across a linear thickness gradient from 2 nm to 12 nm, illustrating the improved resolution of ptychography, especially at greater thicknesses. (C) Ptychography image with enhanced contrast; red arrows mark H atoms. (D) its line profile comparing Pd and PdH$_x$ lattice, showing H incorporation expands the Pd lattice spacing. (E) Depth profile from the same region, mapping the depths of Pd and H columns along the beam direction Z, and highlighting variations in hydrogen occupancy at different depths.

Here, our electron ptychography has 2–3 nm depth resolution along the electron beam (Z) direction, enabling analysis of the 3D structure of PdH$_x$ nanocube. In practice, this means that the nanocube is reconstructed as a stack of 1-nm-thick image slices, resolving atomic structure at successive depths from the top surface to the bottom (Figure 1B). By summing the images in this stack, we generate a 2D projection along Z for comparison between Pd and PdH$_x$, which reveals that hydrogen incorporation leads to lattice expansion (Figure 2D). In addition, by examining the intensity variation along an individual atomic column through the stack (a depth profile), we can observe structural variations along the Z direction: the top and bottom surfaces are slightly inclined (Figure 2E). More importantly, hydrogen is distributed non-uniformly in 3D, with clear hydrogen-depleted regions appearing inside the PdH$_x$ nanocubes.

**1D Hydrogen Superlattice in PdH$_x$ nanocube**

We observed a one-dimensional hydrogen superlattice within PdH$_x$ nanocubes (Figure 3). As noted in the introduction, this is perhaps surprising as hydrogen can diffuse rapidly through Pd at low concentration, but the lattice distortions induced by the hydrogen atoms lead to self-trapping of H, and long-range strain fields can be reduced by clustering, although this has not been reported in the bulk at room temperature.[37] Different diffusion mechanisms have also been noted for bulk and nanoparticle PdH$_x$ where surface strains can also play a role in altering the potential energy surface and may be needed to stabilize some of the self-trapped H into ordered clusters.[50]

We first examined PdH$_x$ nanocubes along the [100] crystallographic orientation (Figure 3A, S9). When the sample is tilted 45° around the y-axis in one direction, corresponding to a [110] viewing orientation, the hydrogen distribution appears homogeneous in projection (Figure 3B, S11). In contrast, tilting 45° in the opposite direction reveals a distinct hydrogen superlattice, characterized by alternating rows of hydrogen and vacancies on the octahedral interstitial sites along the [1$\bar{1}$1] direction (Figure 3C, S12). This geometric relationship implies that the hydrogen superlattice is visible in only three of the six {110} projection directions of PdHx. Therefore, when a nanocube is randomly oriented along a {110} axis, there is a 50% chance of observing the H superlattice in projection. The 1D superlattice of hydrogen atoms within the 3D-symmetric FCC structure of Pd consists of alternating hydrogen planes and vacancy planes.

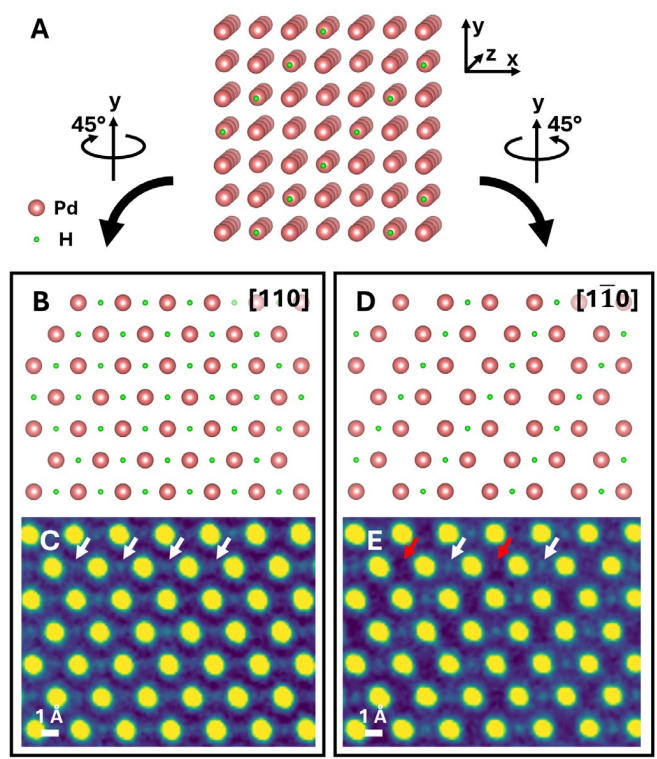

**Figure 3.** (A) Three-dimensional schematic model of PdH$_x$ nanocube based on ptychography results. (B) When the viewing direction is rotated 45° clockwise around the y-axis from [100] to [110], hydrogen atoms appear uniformly distributed in all octahedral interstitial sites. (C) The measured MEP image down this zone shows all interstitial columns occupied with hydrogen (white arrows). (D) When rotated 45° counterclockwise from [100] to [1$\bar{1}$0], a 1D hydrogen superlattice emerges, characterized by alternating hydrogen atom planes. (E) The MEP image shows rows of H planes (white arrows) and vacancy planes (red arrows).

Since hydrogen occupying interstitial sites expands the surrounding Pd lattice, we examined how the hydrogen superlattice affects local Pd lattice spacing (Figure 4). Where a hydrogen superlattice appears, we observe a row-by-row 1D strain wave in the Pd lattice, aligned along the $[1\bar{1}1]$ direction of the superlattice. This pattern consists of alternating tensile (+2%) and compressive (−2%) strain relative to the average lattice spacing of the PdH$_x$ nanocube, as visualized in the strain maps of Figure 4C. This confirms that the ordered hydrogen arrangement directly induces periodic strain modulation in the Pd lattice. The strain modulation propagates from the interior toward the surface and becomes less pronounced near the outermost two atomic layers, where surface-induced strain typically dominates. In contrast, no strain modulation is seen in the direction perpendicular to the hydrogen superlattice, consistent with the absence of ordered hydrogen along this axis (Figure 4D). Similarly, along [110] orientation, where no superlattice is observed, no strain modulation is observed (Figure S13-14).

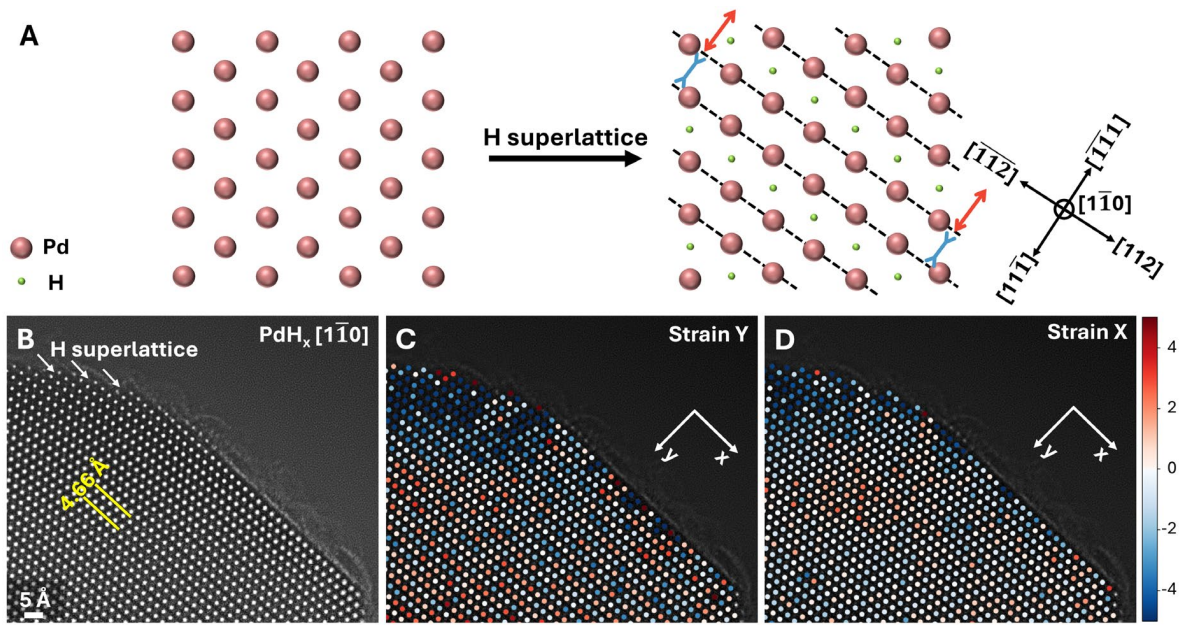

**Figure 4.** (A) Schematic illustration of hydrogen superlattice-induced lattice expansion in PdH$_x$, based on ptychography results. (B) Ptychography image of PdH$_x$ viewed along the $[1\bar{1}0]$ orientation, showing the hydrogen superlattice with a periodicity of 4.66 Å along the $[1\bar{1}1]$ direction. (C–D) Strain maps overlaid on the ptychography image, showing strain modulation wave along the superlattice direction (Y) and the hydrogen rows along perpendicular direction (X), respectively.

## 3D Inhomogeneity of Hydrogen Distribution in PdH$_x$ nanocubes

In addition to the 1D hydrogen superlattice, we observed a 3D inhomogeneous hydrogen distribution within PdH$_x$ nanocubes. The ptychography images reconstruct the nanocube into 1-nm-thick slices along the Z direction, each revealing the positions and aggregation of Pd and H atoms at successive depths (Figure S15-S16). From the imaged region, we assembled a 3D map (Figure 5) revealing that hydrogen is heterogeneously distributed, preferentially accumulating near the nanocube surfaces, while the central interior appears largely hydrogen-depleted.

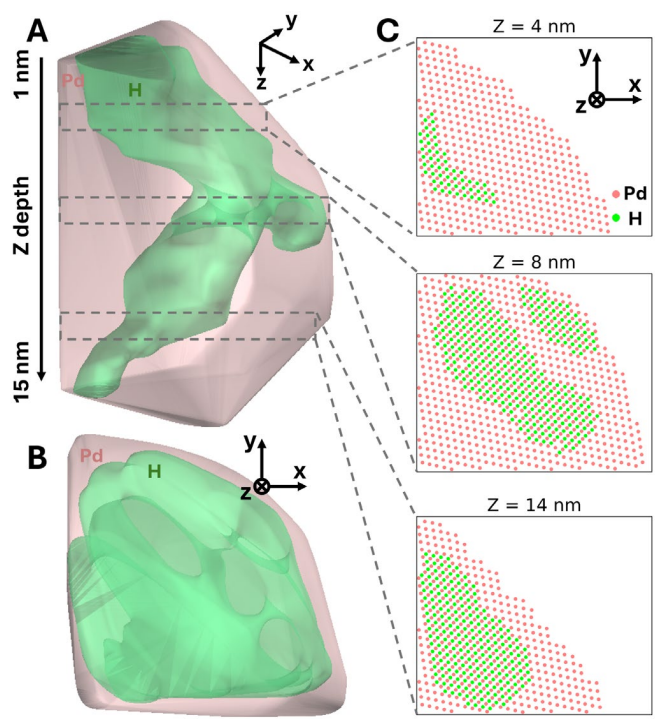

**Figure 5.** (A) Side view and (B) top-down view of 3D distribution of Pd and H atoms in PdH$_x$ nanocube from electron ptychography images. From top-down view, or along the electron beam direction, the H atoms seem to be homogeneously distributed, but from the side view, they are not as indicated at different Z slices. (C) three slices at Z = 4/8/14 nm shows the geometric distribution of H among Pd atoms.

Along the [110] orientation, the reconstructed region exhibits a wedge-shaped geometry, appearing narrower at the top and bottom along the Z axis than at the middle layers. In projection along the beam direction (Z), Pd and H atoms appear uniformly distributed (Figure 5B). However, slice-by-slice analysis reveals that hydrogen aggregates locally within the Pd lattice and gradually shifts laterally along Z, forming a crescent-shaped 3D distribution (Figure 5A, 5C, S17). These hydrogen-rich regions extend approximately 6–7 nm near the surfaces, while the interior remains depleted of hydrogen, suggesting a near-surface energetic preference for hydrogen retention. A similar 3D inhomogeneity is also observed along the $[1\bar{1}0]$ orientation, where the hydrogen superlattice is present at all Z slices (Figure S18), though the depth variation appears less pronounced in that direction.

Finally, we demonstrate an example of the chemical property differences between Pd and $PdH_x$ nanocubes. Electrochemical cyclic voltammetry reveals that Pd nanocubes exhibit a larger hydrogen absorption peak near 0 V vs RHE compared to the $PdH_x$ nanocubes (Figure S19). Differences are also observed in their oxygen reduction reaction (ORR) activity, although this surface-sensitive reaction is beyond the scope of our bulk-structure-focused study. Overall, these results illustrate how structural differences between Pd and $PdH_x$ can lead to distinct chemical properties.

In conclusion, we used electron ptychography to directly image hydrogen atoms in $PdH_x$ nanocubes, revealing both 1D hydrogen superlattices and 3D inhomogeneous distributions, structural insights that were previously inaccessible. The origin of this unexpected hydrogen ordering, its possible presence in other metal hydrides, and its influence on the chemical and physical properties of $PdH_x$ warrant further investigation. More broadly, this work demonstrates electron ptychography as a quantitative technique for directly imaging the 3D distribution of hydrogen atoms with 0.3 Å lateral resolution, opening new opportunities for studies of metal hydrides and related materials.

**Supplementary Figures**

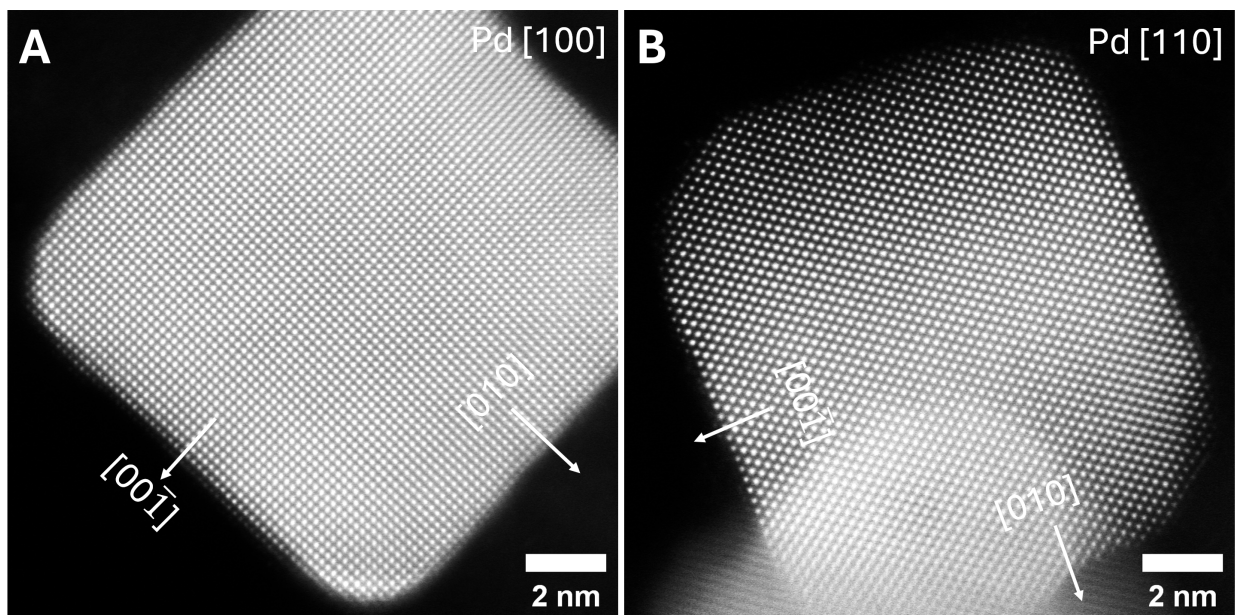

**Figure. S1 Atomic resolution STEM-HAADF images of Pd nanocubes**
(A) Atomic-resolution STEM-HAADF images of Pd nanocubes viewed along the [100] and [110] crystallographic directions, demonstrating their high crystallinity and well-ordered Pd atomic arrangement.

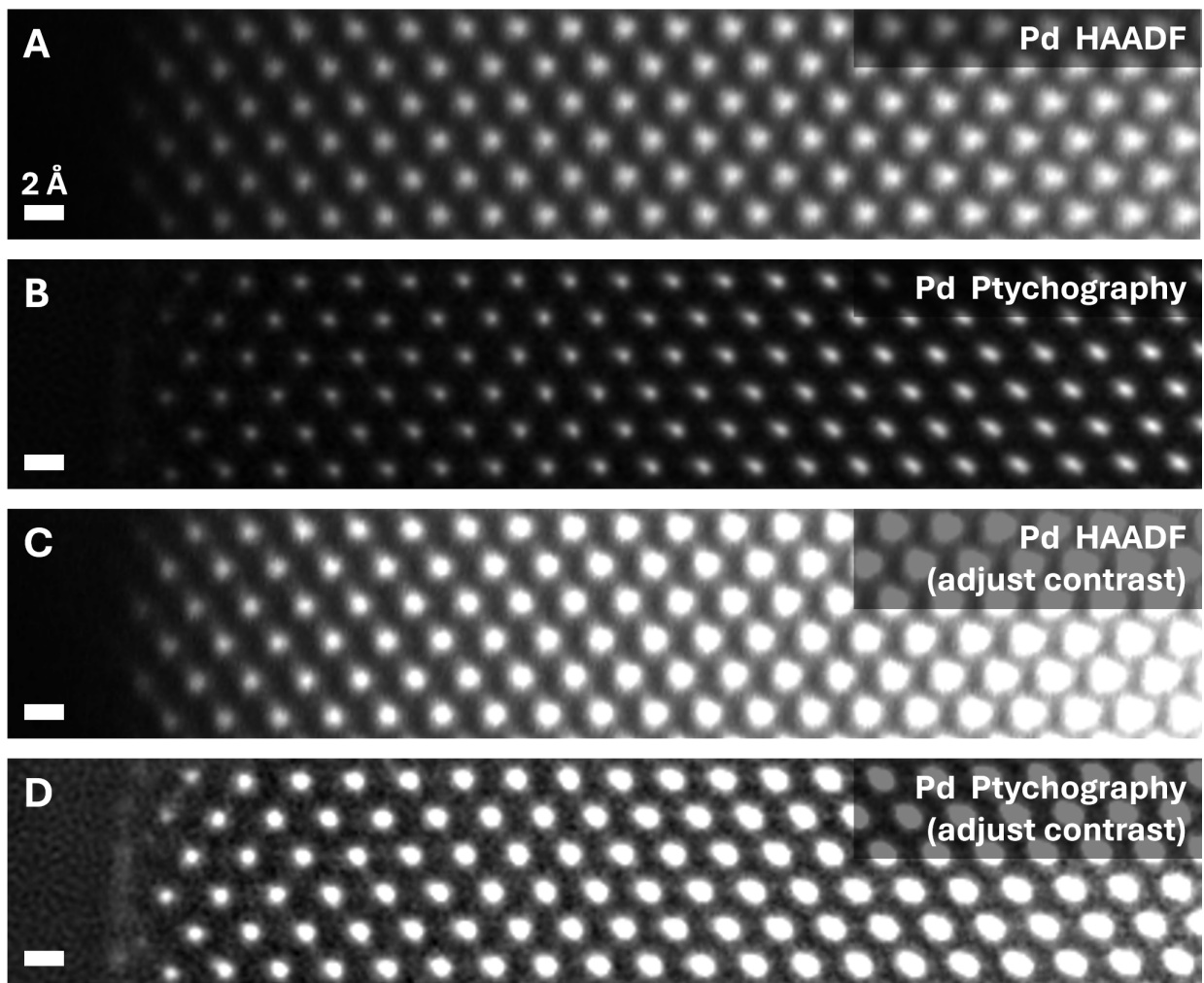

**Figure. S2 Comparison of HAADF Ptychography image of Pd nanocubes along [110]**
(A-B) HAADF and corresponding ptychography images of a Pd nanocube along the [110] direction. Ptychography achieves significantly higher resolution than HAADF, particularly on the right side where the sample is approximately 10 nm thick. The apparent elongation of Pd atoms in the ptychography image is attributed to lateral shifts of Pd columns along the Z direction, as discussed in Figure S3. (C-D) Same images as (A-B) with adjusted contrast. No aggregated hydrogen atoms are observed.

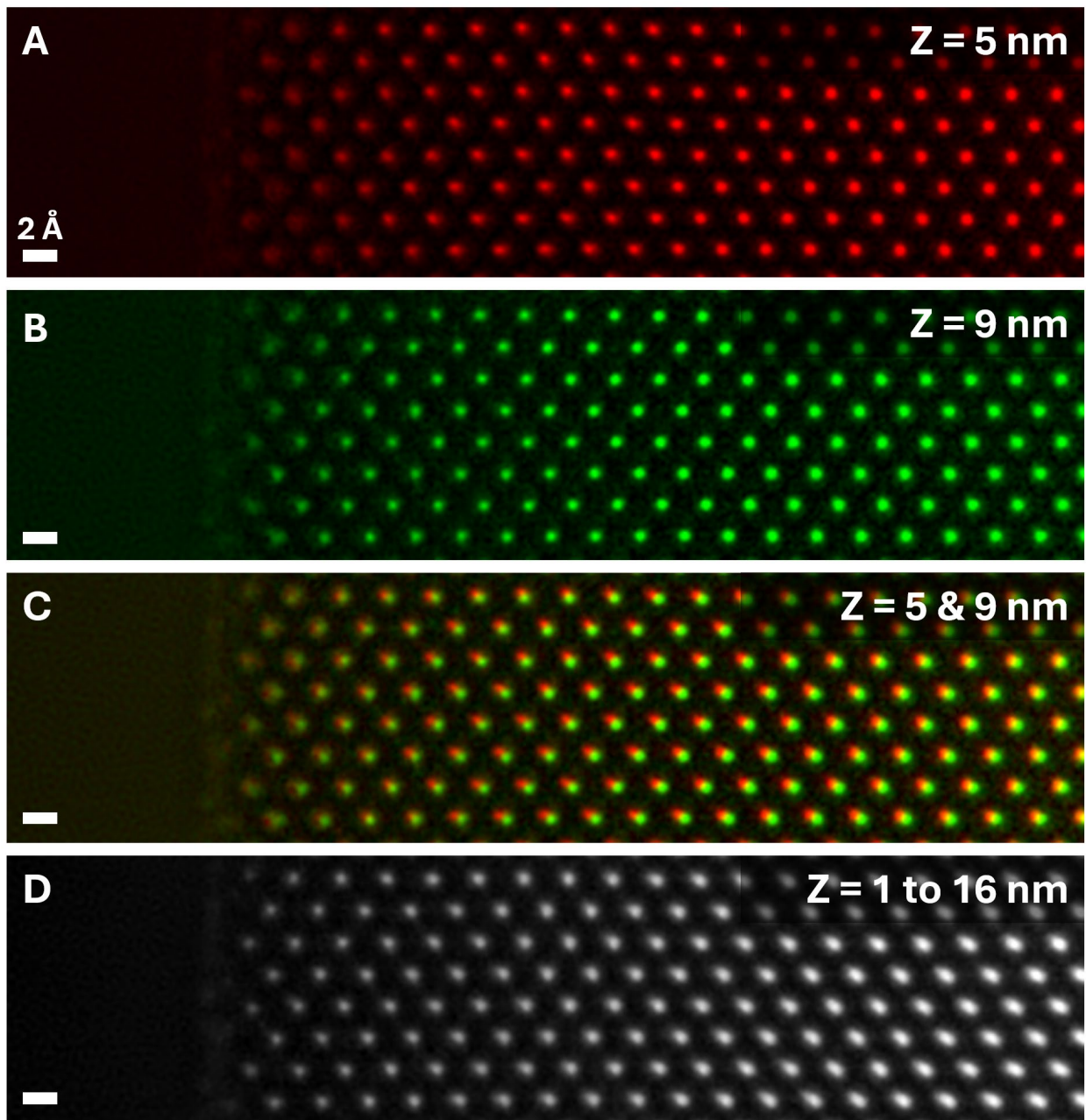

**Figure. S3 Ptychography image of Pd nanocubes along [110] at different Z depth**
Ptychography images reconstruct a 16 nm thick [110]-oriented wedge (including vacuum) of a Pd nanocube into 16 one-nanometer-thick slices. (A, B) Ptychography slices at Z = 5 nm and Z = 9 nm, respectively, showing the atomic structure at different depths. Stacking these slices yields (C), which highlights the lateral shifts of Pd columns along the Z direction. Summing all 16 slices produces (D), where the Pd atoms in thicker regions appear elongated due to these depth-dependent shifts.

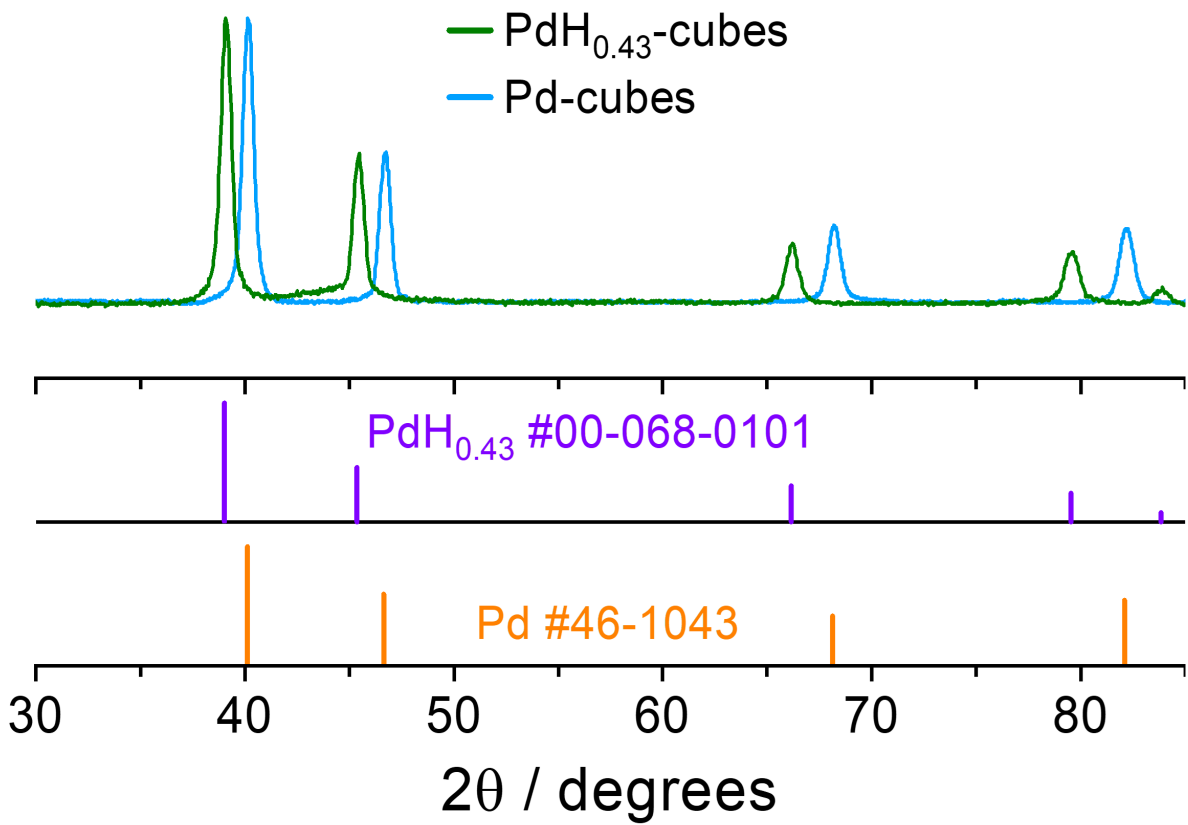

**Figure. S4 XRD patterns of Pd and PdH$_x$ nanocubes**
The XRD patterns of Pd and PdH$_x$ nanocubes match the reference patterns for Pd and PdH$_x$. The incorporation of hydrogen into the Pd interstitial sites expands the lattice parameter, resulting in a characteristic shift of the diffraction peaks to lower angles.

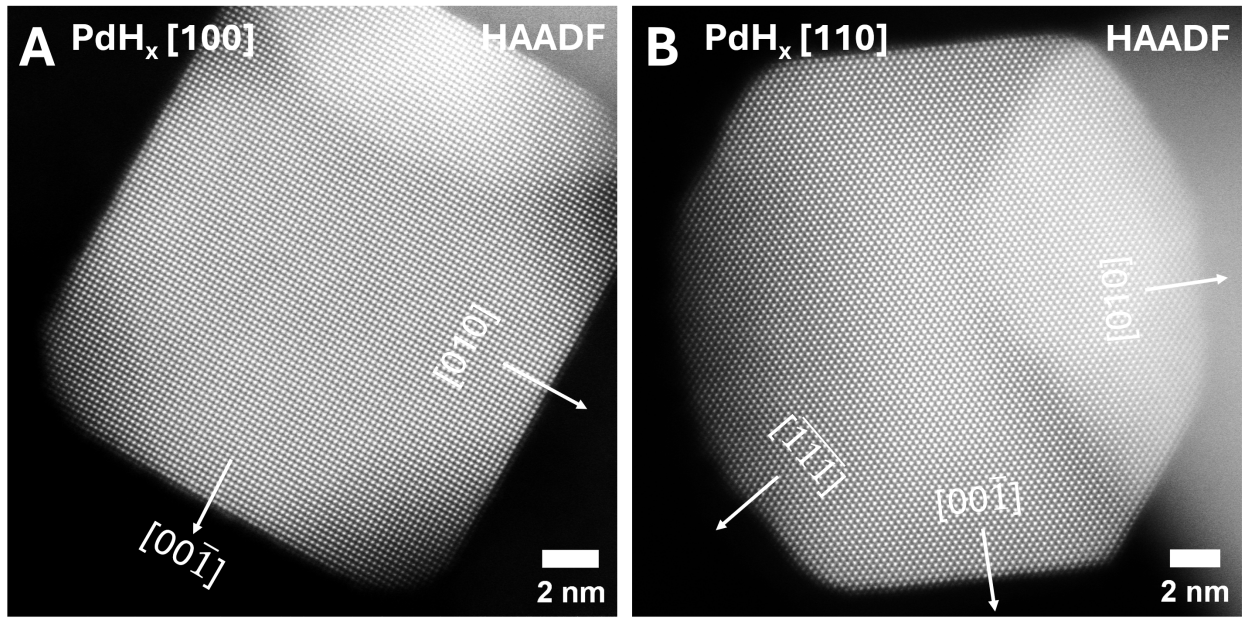

**Figure. S5. Atomic-resolution STEM-HAADF images of PdH$_x$ nanocubes**
(A) Atomic-resolution STEM-HAADF images of PdH$_x$ nanocubes viewed along the [100] and [110] directions, demonstrating their high crystallinity and well-ordered atomic arrangement, similar as Pd nanocubes.

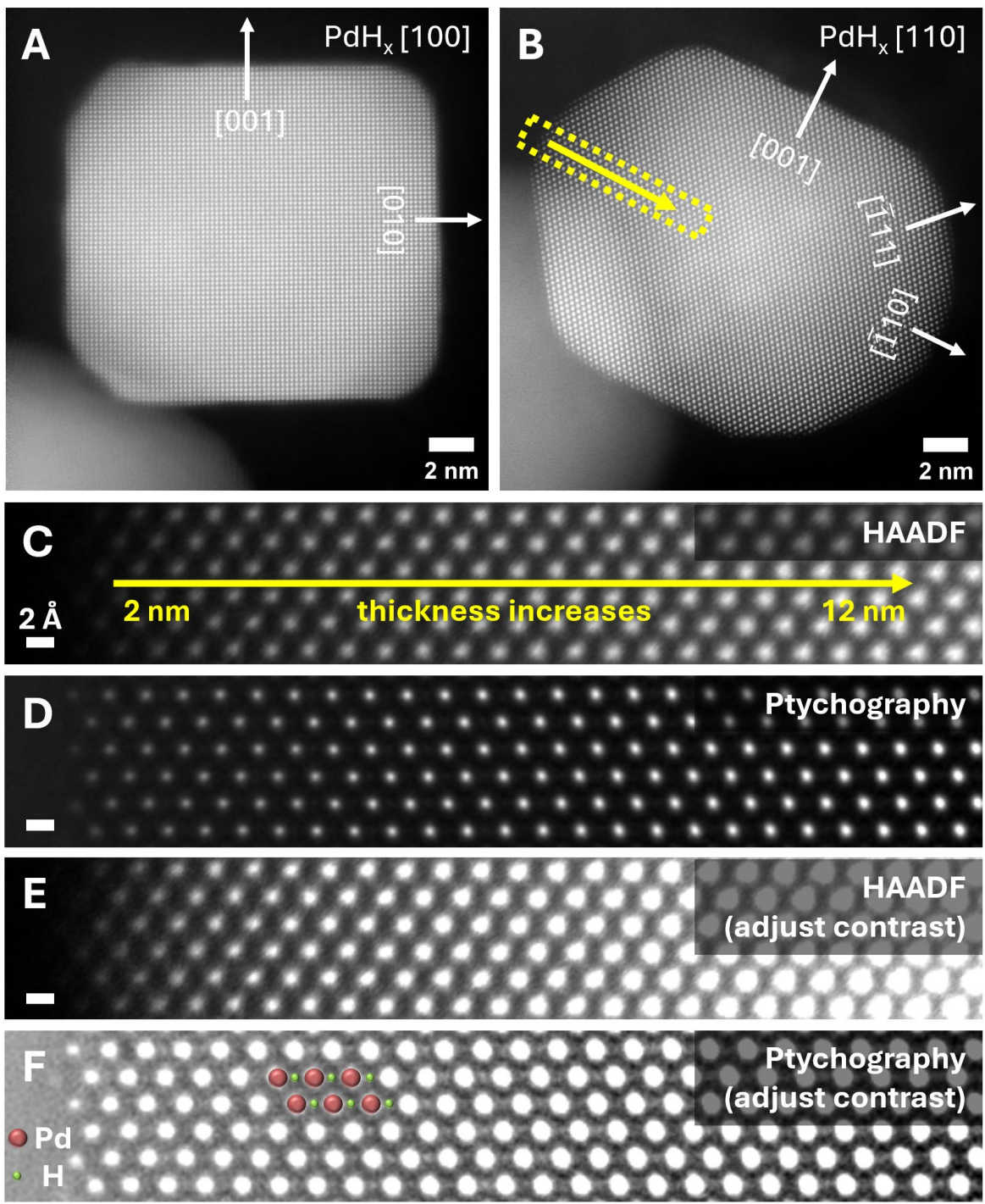

**Figure. S6. Atomic-resolution STEM-HAADF images of PdH$_x$ nanocubes**
(A–B) Atomic-resolution STEM HAADF images of a PdH$_x$ nanocube viewed along the [100] and [110] orientations, respectively. The regions highlighted by yellow boxes are shown in detail in (C–F). (C–D) Comparison of HAADF and ptychography images across a thickness gradient from 2 nm (left) to 12 nm (right), demonstrating that ptychography achieves superior lateral resolution, particularly in thicker regions. (E–F) The same images as (C–D), with adjusted contrast, revealing octahedral hydrogen atoms occupying interstitial sites within the Pd lattice of PdH$_x$

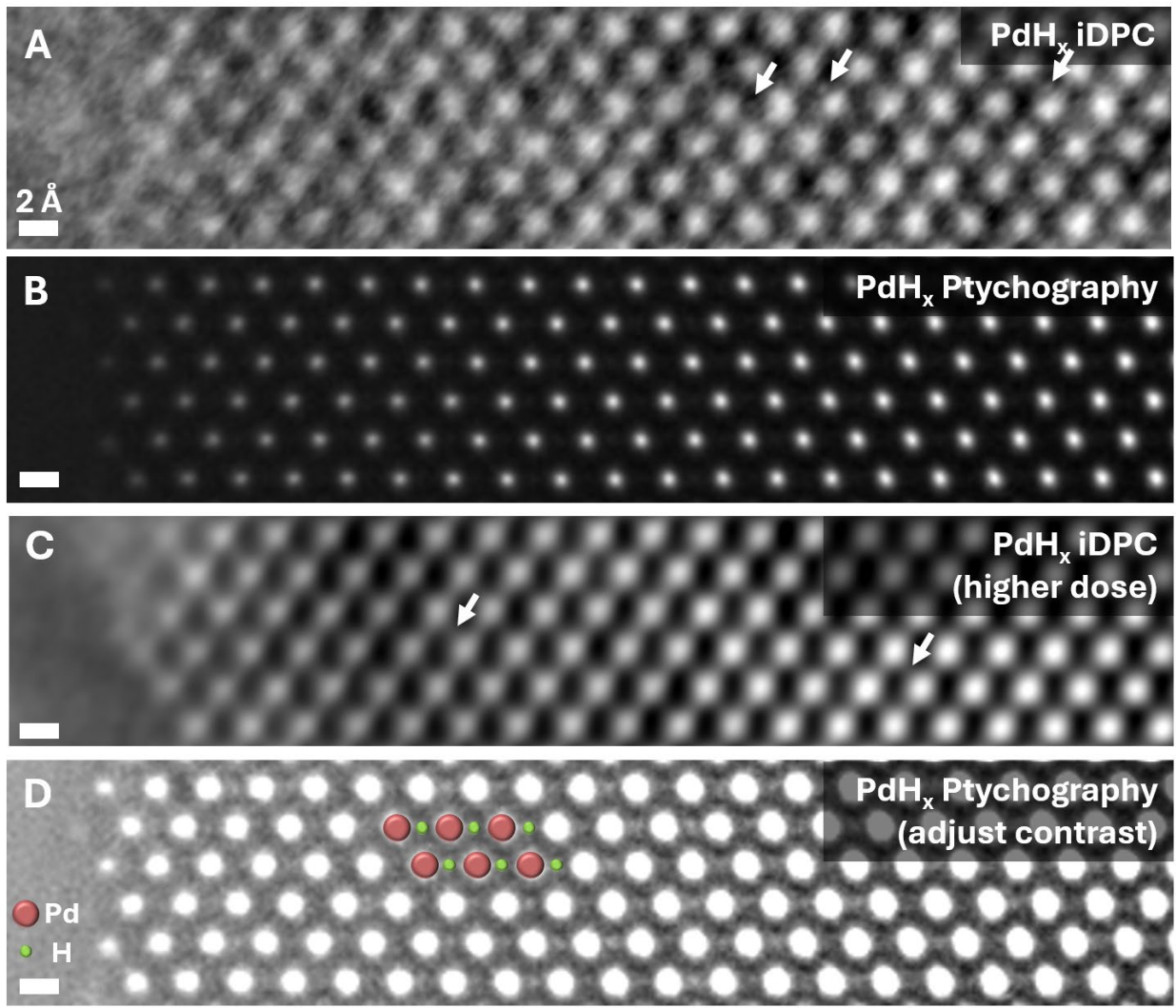

**Figure. S7. Comparison between iDPC and ptychography images of PdH$_x$ nanocubes along [110]**
(A-B) Comparison of iDPC and ptychography images of PdH$_x$ nanocubes along the [110] direction. (C-D) the same images with adjusted contrast. Ptychography accurately locates H atoms, whereas iDPC either fails to resolve hydrogen or incorrectly places them. White arrows indicate apparent hydrogen positions in the iDPC image that, when compared to the ptychography images, are spurious and should not correspond to actual H sites.

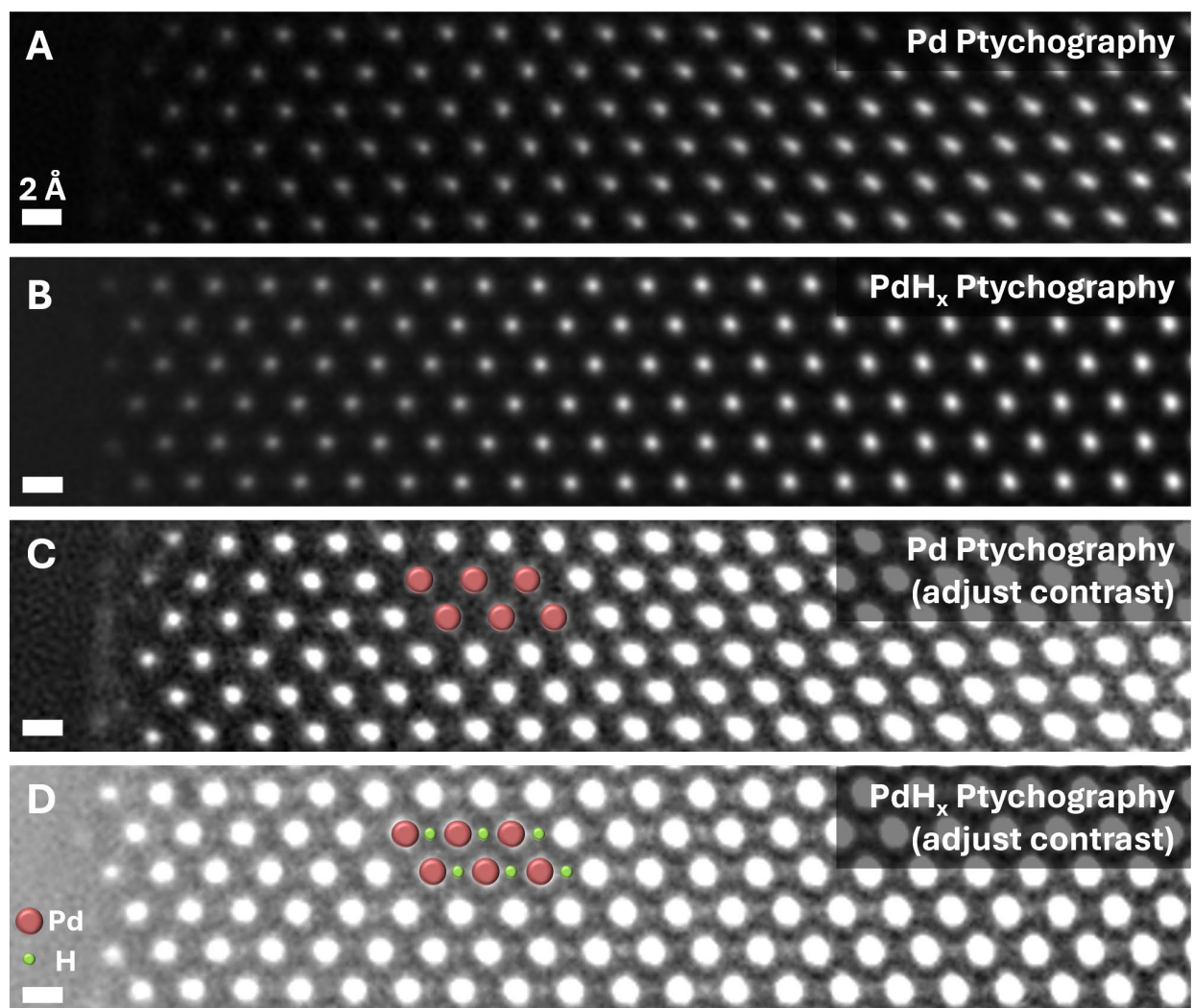

**Figure. S8. Comparison between the ptychography image of Pd and PdH$_x$ nanocubes along [110]**
(A-B) ptychography images of the Pd and PdH$_x$ nanocube, which highlight the Pd atoms. (C-D) The same images with adjusted contrast, allowing observation of both Pd and H atoms. In PdH$_x$, only hydrogen atoms at octahedral interstitial sites are observed; no hydrogen is observed in tetrahedral sites or in Pd.

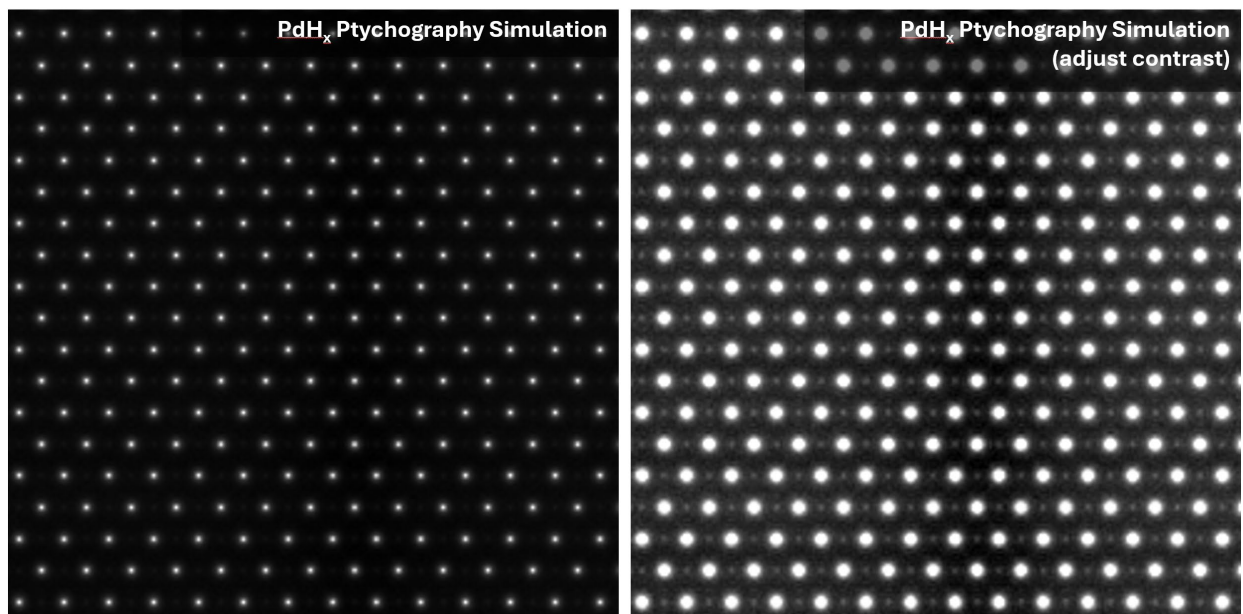

**Figure. S9 Simulated Ptychography image of PdH$_x$ along [1$\bar{1}$0]**
Simulated Ptychography image of 3 nm PdH$_x$ and the same image with adjusted contrast. The simulation condition is the same as the experiment, with less electron dose (100,000 e$^-$/Å$^2$). Just like the experimental data, by adjusting the contrast of the image, the H atoms become clearly visible in simulated data.

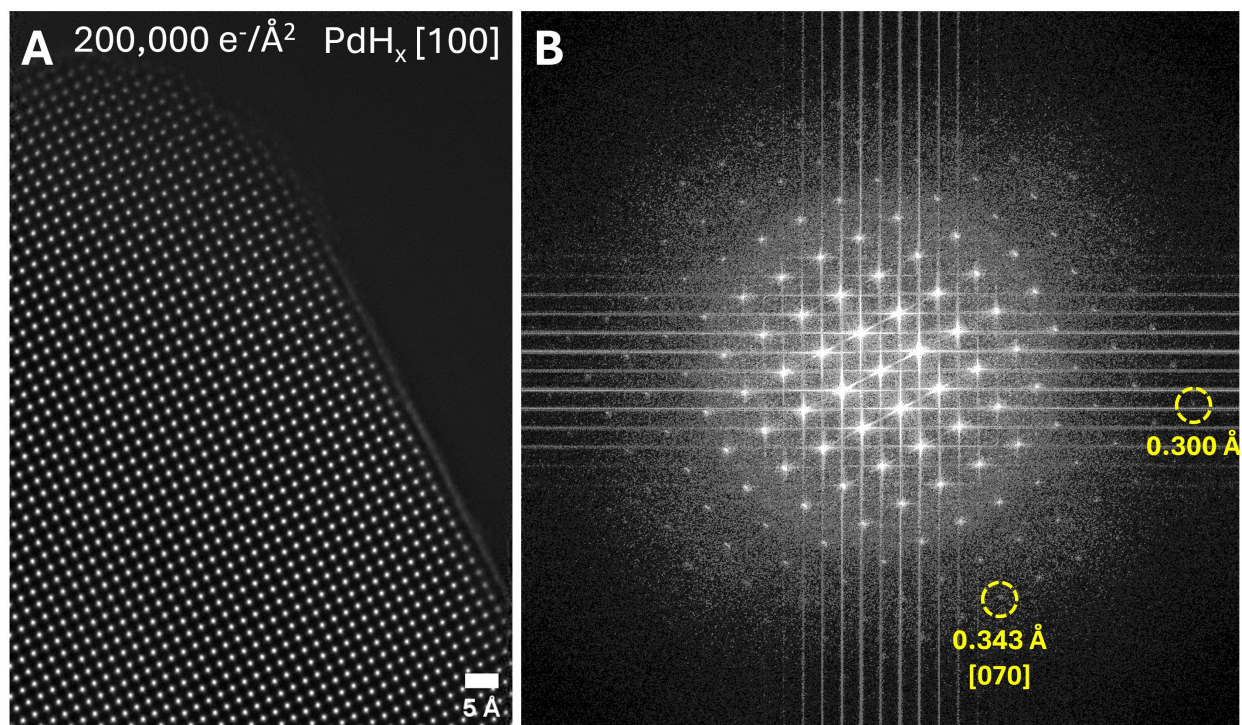

**Figure. S10 Ptychography image of PdH$_x$ nanocube along [100] and its FFT**
(A) ptychography image of PdH$_x$ nanocube along [100], compared to the HAADF image in Figure S4, it has 2 times better resolution, shown by its FFT (B) that the far diffraction spot shows resolution of 0.300 Å with a total electron dose to be 200,000 e⁻/Å$^2$.

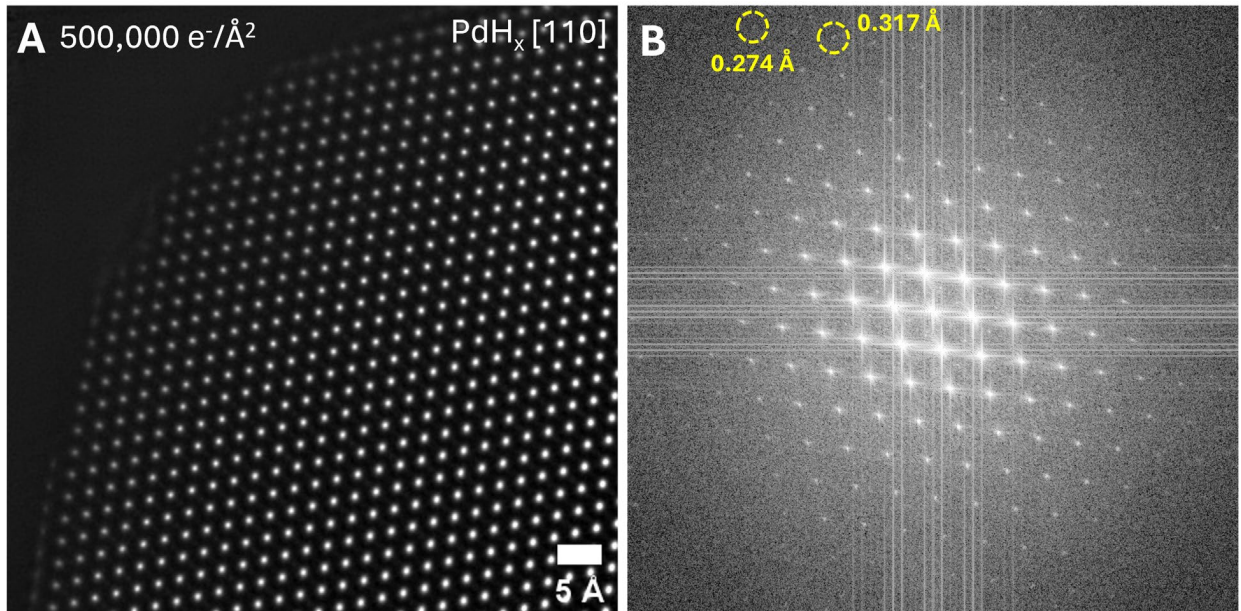

**Figure. S11 Ptychography image of PdH$_x$ nanocube along [110] and its FFT**
(A) ptychography image of PdH$_x$ nanocube along [100], compared to the HAADF image in Figure S4, it has 2 times better resolution, shown by its FFT (B) that the far diffraction spot shows resolution of 0.274 Å with a total electron dose to be 500,000 e$^-$/Å$^2$.

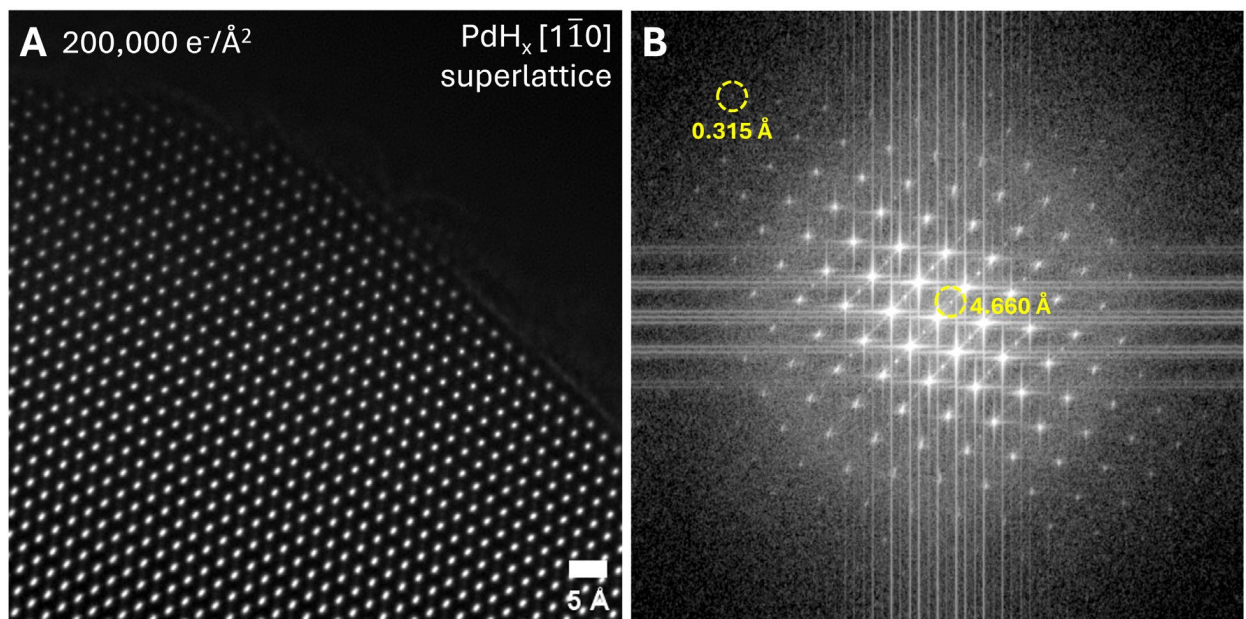

**Figure. S12 Ptychography image of PdH$_x$ nanocube along [1$\bar{1}$0] and its FFT**
(A) ptychography image of PdH$_x$ nanocube along [100], compared to the HAADF image in Figure S4, it has 2 times better resolution, shown by its FFT (B) that the far diffraction spot shows resolution of 0.315 Å with a total electron dose to be 200,000 e$^-$/Å$^2$. The superlattice peak 4.660 Å, which is only along one direction, is clear, but is not observed along the [110] orientation.

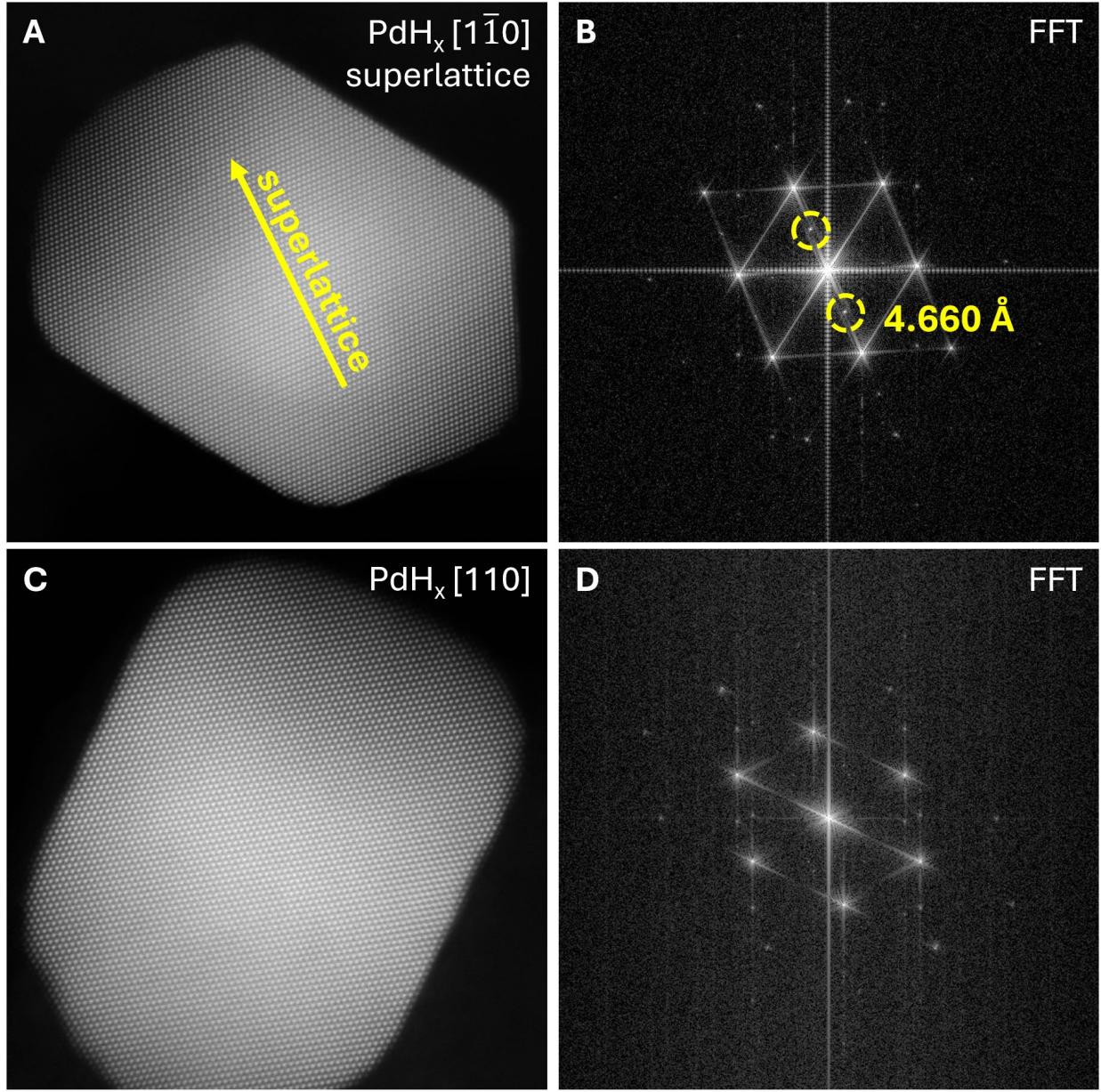

**Figure. S13 HAADF image of PdH$_x$ nanocube along [1$\bar{1}$0] and [110] and their FFT**
(A-B) HAADF image of PdH$_x$ nanocube along [1$\bar{1}$0] and its FFT showing the 1D superlattice. Noted: The H superlattice induced Pd strain wave can be observed throughout the whole PdH$_x$ nanocube. (C-D) HAADF image of PdH$_x$ nanocube along [110] and its FFT showing the 1D superlattice, no superlattice is observed.

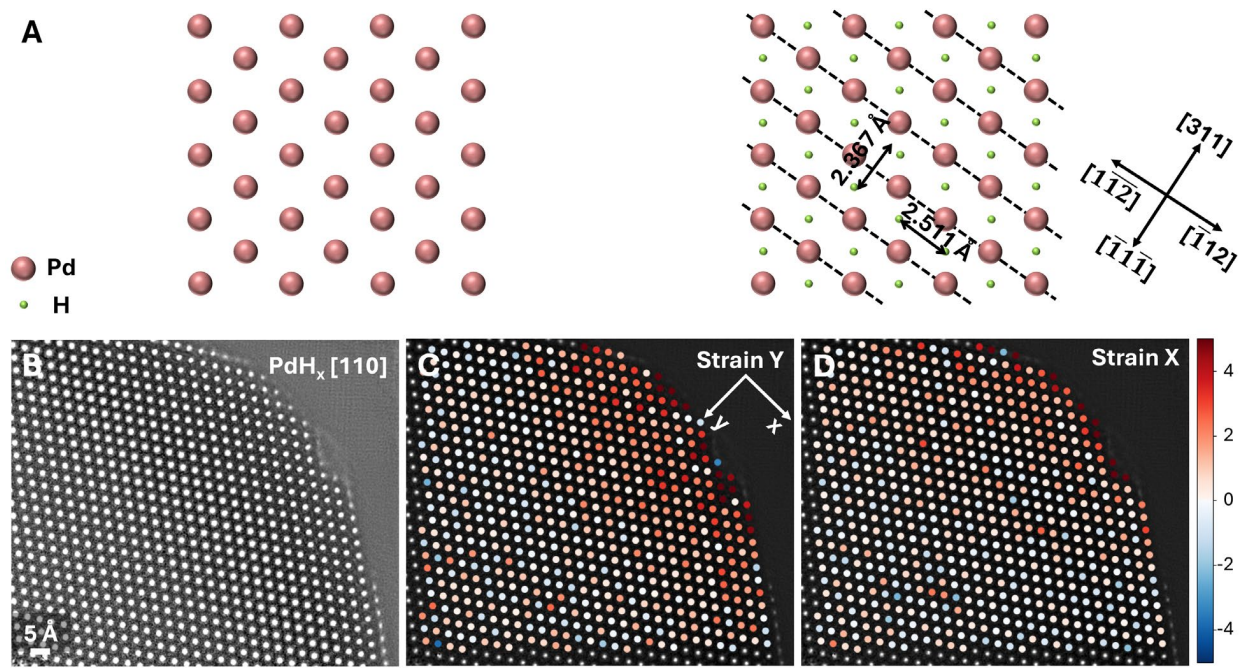

**Figure. S14 Strain map of PdH$_x$ along [110] with H homogenously distributed**
(A) Schematic illustration of lattice expansion induced by a hydrogen superlattice in PdH$_x$, based on ptychography results. (B) Ptychography image of PdH$_x$ showing hydrogen atoms homogeneously distributed within the Pd lattice along the [$\bar{1}1\bar{1}$] direction. (C-D) Strain maps extracted from the ptychography image along the H superlattice direction and the hydrogen atom row direction, showing no periodic strain modulation, contrasting with the strain maps observed for regions with an H superlattice.

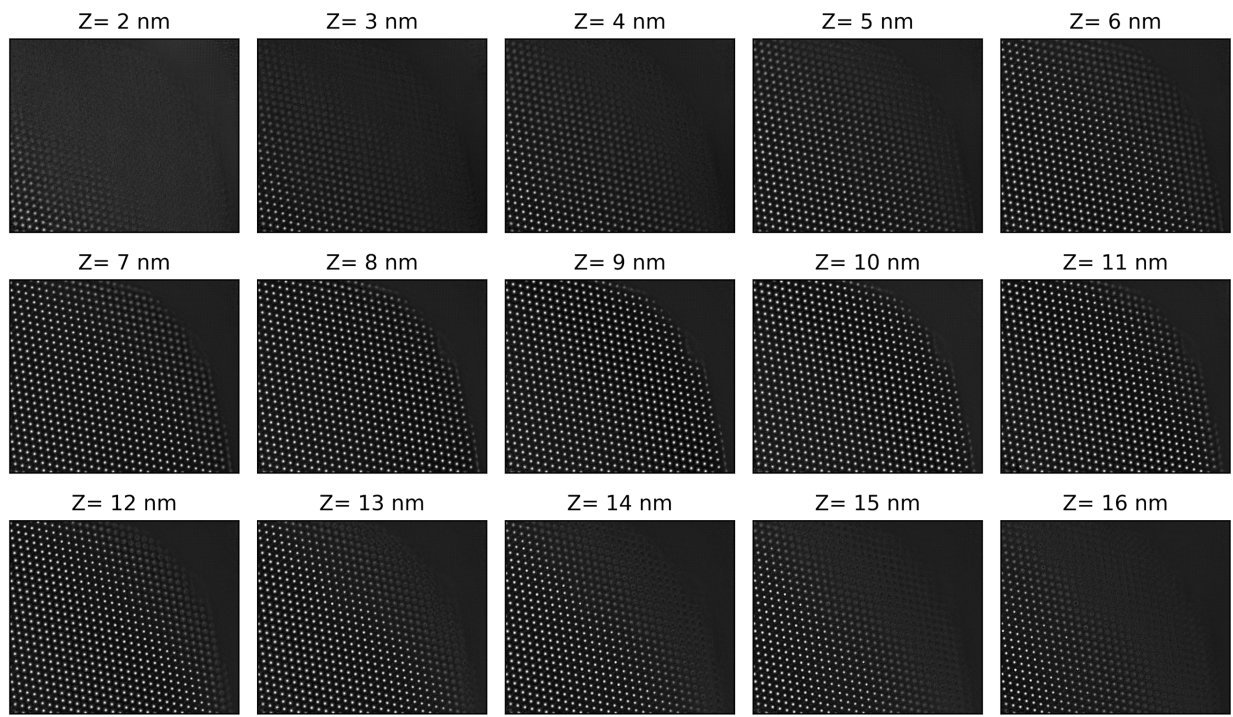

**Figure. S15 Fifteen 1-nm slices of ptychography images for PdH$_x$ nanocubes at different Z depths along [110]**

This ptychography image stack reveals the wedge-shaped geometry of the PdH$_x$ nanocube along the [110] axis. From these slices, the positions of Pd and H atoms can be determined in each layer, allowing reconstruction of the 3D distribution of Pd and H within the nanocube.

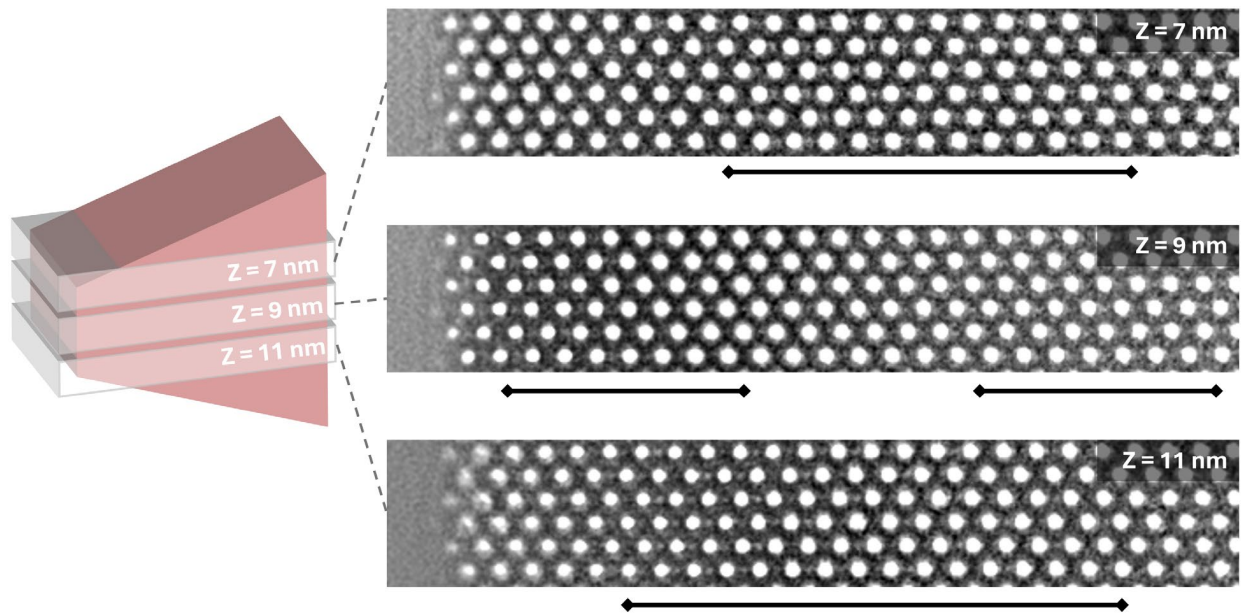

**Figure. S16 H distribution at different Z depths of ptychography images**
This ptychography image stack from one reconstruction dataset reveals the hydrogen distribution at different Z depths. The left is a schematic model of the $PdH_x$ nanocube along [110] orientation and it's a wedge shape. The right shows different 1-nm thick slices, with the clustering of H atoms at specific regions at highlighted in black lines.

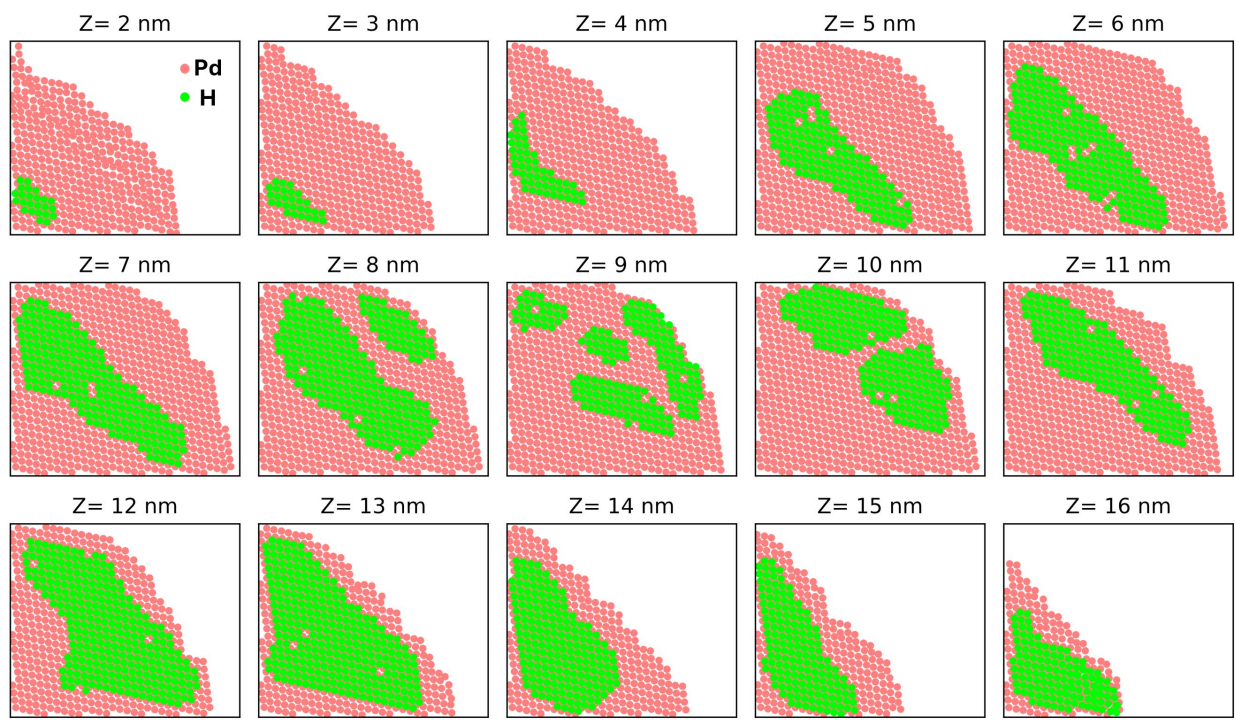

**Figure. S17 Distribution of H and Pd atoms at different Z depths along [110]**

Pd and H atoms are identified on each ptychography slice. While individual Pd atoms can be precisely located, hydrogen is detected as regions of aggregation rather than as isolated atoms. Using the 2D distributions from slices at depths ranging from 2 nm to 16 nm, a 3D hydrogen distribution within the Pd lattice is reconstructed by applying a shape interpolation/deformation method to produce a smooth, continuous 3D representation. These aggregated H regions exhibit a gradual deformation and shift between adjacent slices along the Z axis.

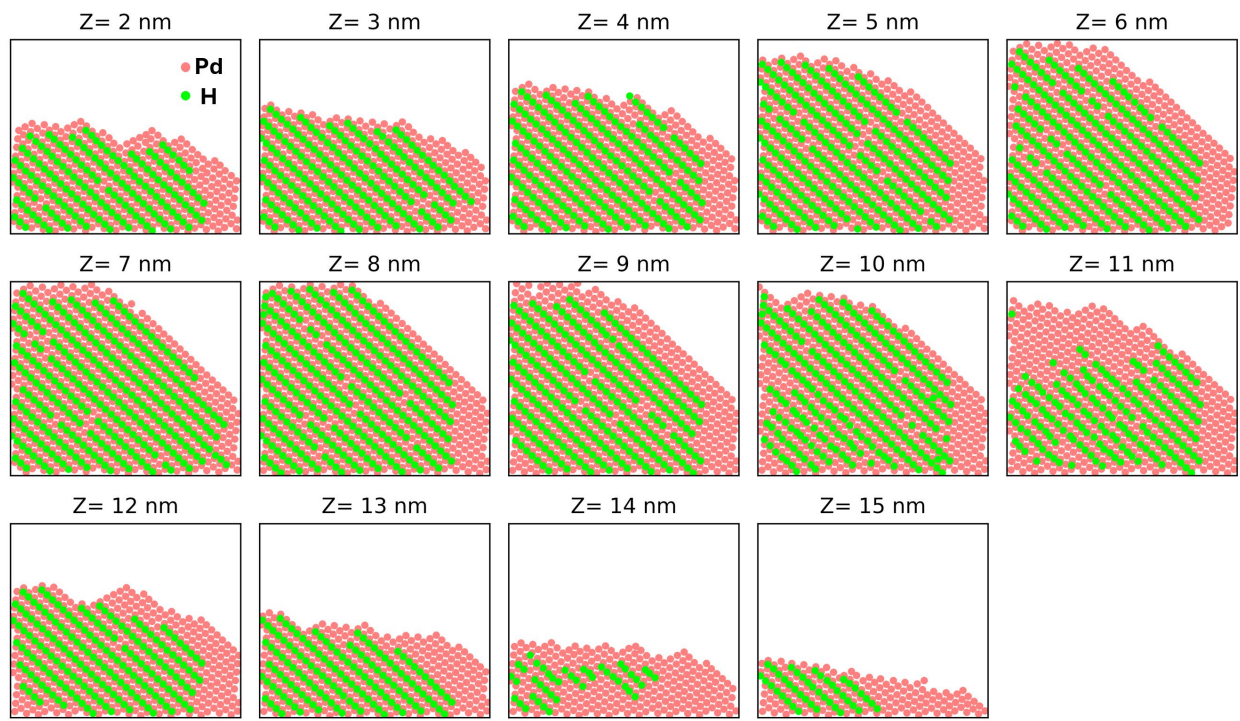

**Figure. S18 Distribution of H and Pd atoms at different Z depths along [1$\bar{1}$0]**
The ptychography image stack along [1$\bar{1}$0] reveals that a hydrogen superlattice is present on each slice at different depths. However, the hydrogen distribution remains inhomogeneous, both within individual 2D slices and in the overall 3D reconstruction.

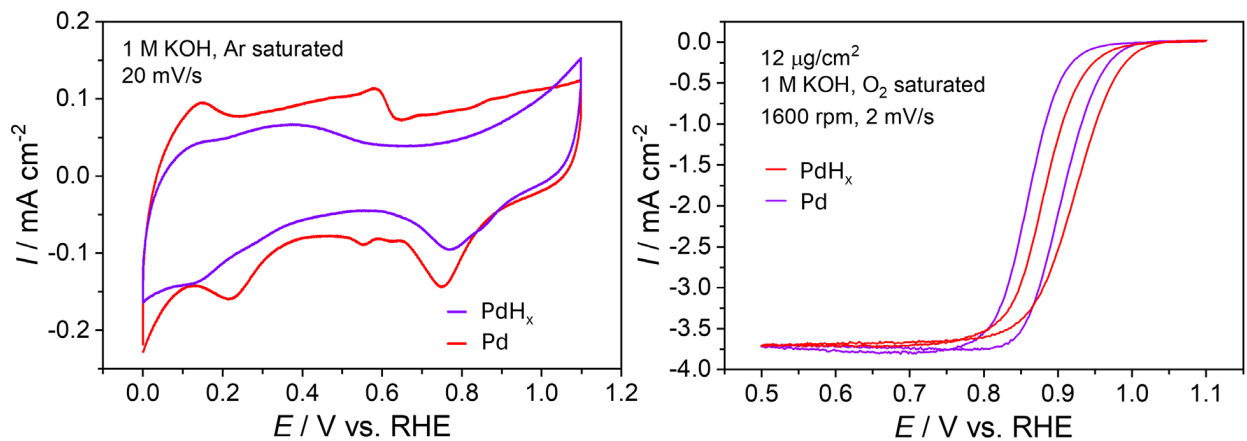

**Figure. S19 Electrochemical measurement of Pd and PdH$_x$ nanocubes**
The cyclic voltammetry (CV) (left) and oxygen reduction reaction (ORR) (right) profiles comparison between Pd and PdH$_x$ nanocubes.